\documentclass[preprintnumbers,prc,twocolumn]{revtex4}
\usepackage{graphics}
\usepackage{bm}
\usepackage{hyperref}
\usepackage{xcolor}
\usepackage{graphicx} 
\usepackage{subfigure}
\newcommand{\lm}{\ensuremath{\Lambda} }
\newcommand{\antilm}{\ensuremath{\bar{\Lambda}} }

\newcommand{\tev}{\ensuremath{\sqrt{s} = } }
\newcommand{\seventev}{\ensuremath{\sqrt{s} = 7} TeV }

\newcommand{\pt}{\ensuremath{p_{\rm{T}}} }
\newcommand{\mody}{\ensuremath{|y|} }

\newcommand{\namepythia}{\ensuremath{\textsc{P}\textsc{ythia}} }
\newcommand{\pythiamode}{\ensuremath{\textsc{P}\textsc{ythia\_}}}

\begin{document}

\title{Baryon number transportation over full rapidity space in pp collisions at LHC energies}
\author{ Banajit Barman\textsuperscript{1}\footnote{banajit.barman@cern.ch}}
\author{ Nur Hussain\textsuperscript{1,2}\footnote{nur.hussain@cern.ch}}
\author{Buddhadeb Bhattacharjee\textsuperscript{1}\footnote{buddhadeb.bhattacharjee@cern.ch (corresponding author)}}
\affiliation{ \textsuperscript{1}Nuclear and Radiation Physics Research Laboratory, Department of Physics, Gauhati University, Guwahati, 781014, Assam, India \\
\textsuperscript{2}Department of Physics, MC College, Barpeta, 781301, Assam, India }

\begin{abstract}
  The estimation of anti-baryon to baryon ratio is considered to be a useful tool for studying baryon number transport in pp, pA and AA collisions. For this study, the \namepythia event 
  generator with various tunes is used to measure the $\bar{\Lambda}/ \Lambda$ ratio as a function of rapidity (y), transverse momentum (\pt), and multiplicity within both ALICE and LHCb acceptances. 
  The results obtained using various MC data for pp collisions at $\sqrt{s} = 7$ TeV are compared with the experimentally measured values of $\bar{\Lambda}/ \Lambda$ ratio
  of ALICE and LHCb experiments. Out of the various studied tunes of \namepythia 8.3 the string junction model is found to be most successful to describe the experimentally 
  observed results. Evidences of a considerable amount of baryon number transportation from the beam fragmentation to the ALICE and LHCb  acceptances could be recognised.
\end{abstract}
\date{\today}
\maketitle

\section{Introduction}

In heavy ion collisions, the antiparticle to particle ratio of hadrons have been extensively studied at various RHIC and LHC energies 
\cite{back2001ratios,adler2003midrapidity,tawfik2015particle,abelev2013centrality}. To have an insight into the mechanism of baryon number transport and to 
evaluate available 
energy from the collision for particle production, such an estimation of antiparticle to particle ratio is found to be of considerable significance \cite{bearden2004nuclear}. 
Besides these, information on net baryon density and baryon chemical potential can also be extracted from such a measurement on \antilm to \lm ratio \cite{braun1999chemical}.  
\par 
In the framework of Quantum Chromodynamics(QCD), generally the valence
quarks carry the properties such as color, flavor, electric charge, and isospin of the baryons. However, Garvey et al. in reference \cite{garvey2000baryon} has pointed out with the example of strong interaction $\pi^{-} + p \rightarrow \Omega^{-} + K^+ + 2K^0$ that the assignment
of baryon number to specific parton of a baryon is not straightforward as $\Omega^{-}$ of the above equation does not contain any valence quark
of the proton, although the baryon number is conserved. This leads them to propose that there may be
other partons responsible for carrying the baryon number of the proton. References
~\cite{garvey2000baryon,kharzeev1996can,kopeliovich1999baryon} suggest that one such parton capable of carrying the baryon number 
over a large rapidity range is the gluon or a gluonic string junction. A gluonic string junction is considered to be a single point in QCD 
at which three valance quarks join at that point. In this picture the baryon number is associated with gluonic field \cite{alice2013mid}.
Further, in ref.~\cite{kharzeev1996can}, Kharzev has pointed out that the gauge 
invariance constraint of a strongly interacting physical particle of QCD requires that 
the trace of the baryon number should be associated not with the valance quark but with 
the gluon fields located at the string junction.
Hence, there remains ambiguity regarding whether to attribute the baryon number to the valence 
quarks or to the gluonic field. To gain insight into this conundrum, and to understand baryon transport mechanism, one needs to examine the mechanisms of baryon and anti-baryon production in 
pp, pA, and AA collisions over full rapidity space. 
\par
In baryon-baryon collisions, there are two probable mechanisms for baryon production and one for anti-baryon production. One of the mechanisms 
of baryon production
involves pair production from the vacuum via string and anti-string junction pair production, followed by their combination 
with sea quarks and antiquarks. The other mechanism entails baryon production from constituents of the beam, where the baryon may 
originate from valence quarks, diquarks, or the string junction (or sometimes from a combination of these three) \cite{alice2013mid}. 
The spectra of the 
baryon can differ from that of anti-baryon at mid-rapidity if any of the constituents get diffused over large rapidity interval. It therefore becomes very much pertinent to experimentally measure the ratio of 
anti-baryons to baryons across a wide rapidity range and compare the results with the prediction of various MC models. Unfortunately, the existing detectors of LHC are capable of measuring the ratio only over limited acceptance.
While the ALICE collaboration measured \antilm / \lm ratio in pp collisions 
at LHC RUN 2 energies \tev $0.9, 2.76$, and $7$ TeV as a function of rapidity (y) and transverse 
momentum (\pt) within the acceptance \mody $ <$ 0.8 \cite{alice2013mid}, LHCb collaboration measured the same parameters 
for pp collision at \tev 0.9 and 7 TeV within the rapidity range ($2<y< 4.5$) \cite{aaij2011measurement}.
LHCb collaboration, by comparing the ratio with various 
tunes of \textsc{P}\textsc{ythia}6, such as LHCb MC, Perugia 0, and Perugia NOCR, observed that none of these \textsc{P}\textsc{ythia}6 tunes could 
adequately explain the data of LHCb across the rapidity range of $2 < y < 4.5$. However, the \namepythia tunes are being constantly improvised to obtain
a better agreement between various experimentally observed results and the model prediction, which ultimately resulted in a completely new version of \textsc{P}\textsc{ythia}, namely  \textsc{P}\textsc{ythia}8.
Besides the upgradation of \textsc{P}\textsc{ythia} model, LHC has also upgraded its experiments time to time from RUN1 to RUN2 and subsequently to RUN3.
Accordingly, \textsc{P}\textsc{ythia} 8 has been tuned by introducing more \& more underlying processes of particle production and transport
mechanisms to match with the experimental results of ALICE and LHCb experiments for LHC RUN 2 and RUN 3 data. So, it would be interesting to compare the existing
ALICE and LHCb experimental results on rapidity and \pt distributions with the various tunes of latest version of \namepythia 8 and look for a single \namepythia
MC event generator that incorporates more realistic pp collision dynamics to agree with the ALICE and LHCb RUN 2 results on the above-mentioned global
observables and predict \antilm / \lm ratio over full rapidity space for LHC RUN 3 data.

\section{Event Generator}

Towards the end of last century, for the study of pp collisions, \textsc{P}\textsc{ythia} \cite{bengtsson1984lund,bengtsson1984improved,bengtsson1987lund} and  \textsc{J}\textsc{etset} \cite{sjostrand1987lund} models were widely utilized for event generation.
Both of these models had fragmentation as process for hadronization. Later, \textsc{P}\textsc{ythia} 5.7 and \textsc{J}\textsc{etset} 7.4 were merged, 
and a new version of \textsc{P}\textsc{ythia}, namely \textsc{P}\textsc{ythia} 6, was released. \textsc{P}\textsc{ythia}6 had mainly four releases from 
\textsc{P}\textsc{ythia} 6.1 to 6.4. 
With each release, the physics content of the model was updated. After merging, the hadronization process in \textsc{P}\textsc{ythia} 6 was 
solely based on Lund String Fragmentation framework. In \textsc{P}\textsc{ythia} 6, along with some minor changes to the old popcorn mechanism \cite{andersson1982model,andersson1985baryon}, a new popcorn mechanism scheme for 
baryon production was introduced \cite{eden1997baryon,eden1996program}.  

In comparison to old popcorn mechanism that permitted at most one meson to produce between a pair of baryon and anti-baryon, the new model allowed an arbitrary 
number of mesons to produce. Further, in this version color rearrangement and alternative Bose-Einstein description, which represents correlation between two identical bosons such as pions etc., were incorporated \cite{sjostrand2001high}. 
The string hadronization scheme was also improvised by including the junction topologies, where three strings meet and was more
suitable in specific scenarios, such as when baryon number was violated or when more than one valence quark was kicked from the incoming proton. Moreover, a new framework for multiple interactions was 
introduced for the study of minimum-bias events or events underlying hard process in hadron-hadron collisions as mentioned in reference \cite{sjostrand2006pythia}.The more details about the 
junction topology in \textsc{P}\textsc{ythia} can be found in reference \cite{sjostrand2003baryon}. Along with these updates in physics processes, several tunes as mentioned in reference \cite{skands2010tuning} 
were incorporated in \textsc{P}\textsc{ythia}6. Some of them were Perugia 0, Perugia HARD, Perugia SOFT, Perugia NOCR, Perugia 6 etc. These tunes were based on different values of 
parameters related to final state radiation, underlying-event, beam remnant, colour reconnection and hadronization. Every tune had its own physics motive. For instance, the Perugia NOCR tune was introduced to include a large amount of baryon number transport, utilising 
significant 'color disturbance' in the remnant. These tunes of \textsc{P}\textsc{ythia} 6 
were used to compare with results of RUN 1 of the LHC and can be found in reference \cite{katzy2013qcd,sjostrand2008brief,corke2011interleaved,atlas2010charged,maestre2012sm,firdous2013tuning}. However, 
\textsc{P}\textsc{ythia}6 is found to be inadequate to describe a number of experimental results of the LHC RUN 2 data necessitating further improvement and hence yet another 
upgraded version, which is \textsc{P}\textsc{ythia}8, was made available.
\par 
\textsc{P}\textsc{ythia}8 serves as the successor to \textsc{P}\textsc{ythia}6, incorporating numerous changes, which currently have three releases namely 8.1, 8.2 and 8.3. However, \textsc{P}\textsc{ythia }8.1 model, to some extent was unsuccessful 
in explaining a number of results of the LHC RUN 2 data. \namepythia 8.2 is a more matured version of \textsc{P}\textsc{ythia}8.1 that incorporates several updates in different processes like parton shower, multiparton interactions, beam remnant and color reconnection. 
Even though the hadronization process in 8.2  still  solely based on Lund String Fragmentation framework, few additional improvements were made such as the handling 
of junction topologies which now allow more complicated multijunction string configuration \cite{sjostrand2015introduction}. In addition to the tune 4C (RUN 1), in \namepythia 8.2, a new tune named as 
“Monash” that covers LEP, Tevatron and LHC data was introduced in the year 2013 \cite{skands2014tuning}. Later, this tune was set as default tune in 
\textsc{P}\textsc{ythia}8.2. With this tune, the experimental results of p+p collisions at  LHC RUN2 energies were found to be agreed well \cite{skands2014tuning}. However, there is a visible disagreement between the 
\namepythia 8.2 prediction and experimentally measured values of a number of observables of collective like behaviours such as $<\it{p}_T>$ vs multiplicity, $v_{2}$ vs \pt etc \cite{acharya2019multiplicity,alice2017enhanced}. Hence, latest version of \namepythia 8 i.e. \namepythia 8.3,
that incorporates string shoving mechanism, has also been developed \cite{bierlich2022comprehensive}. Generally, in the Lund string model, strings are considered as massless relativistic strings and they do not have transverse extensions. 
However, 
this approximation does not hold in a scenario with many strings occupying the available spatial volume, as in the case of hadronic collisions. Hence, 
in such a scenario, the strings are allowed to interact mainly through repulsive forces. This forms the core concept behind the string shoving model. 
In \textsc{P}\textsc{ythia}, the model does not presume the presence of a thermalised medium like the Quark-Gluon Plasma (QGP). This indicates that \textsc{P}\textsc{ythia}
can not presume such observable which are traditionally attributed to Quark-Gluon Plasma (QGP) formation \cite{bierlich2022comprehensive}. Hence, \textsc{P}\textsc{ythia} extended the Lund string model,
allowing the interactions between strings in densely populated regions of space to explain the collective effect observed  particularly in proton-proton collisions \cite{bierlich2022comprehensive}. In other words, the string 
shoving model has been introduced to mimic the observables of collective motion in a QGP like scenario. Currently \textsc{P}\textsc{ythia}8 includes many underlying events 
in Lund String Fragmentation Model like color 
reconnection, string shoving, rope hadronization to explain 
collective behaviour in pp collision \cite{bierlich2024string}. More details about the current versions of \textsc{P}\textsc{ythia}8 can be found in ref \cite{bierlich2022comprehensive}.

\section{Results and discussions}

\begin{figure*}
  \subfigure[]{
    \includegraphics[width=0.45\linewidth]{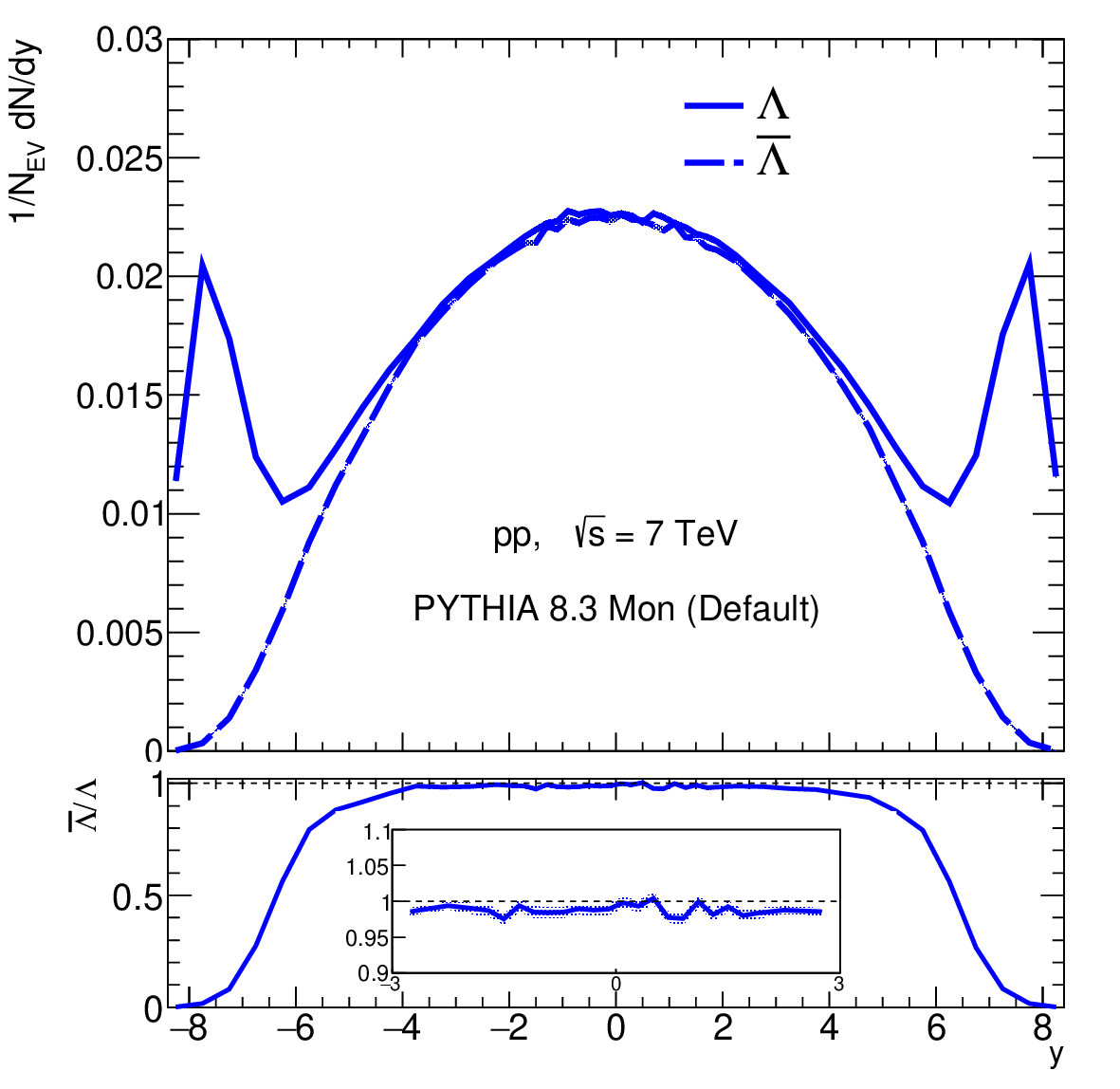}
    \label{fig:mon_rapidity}
    
  }
  \subfigure[]{
    \includegraphics[width=0.45\linewidth]{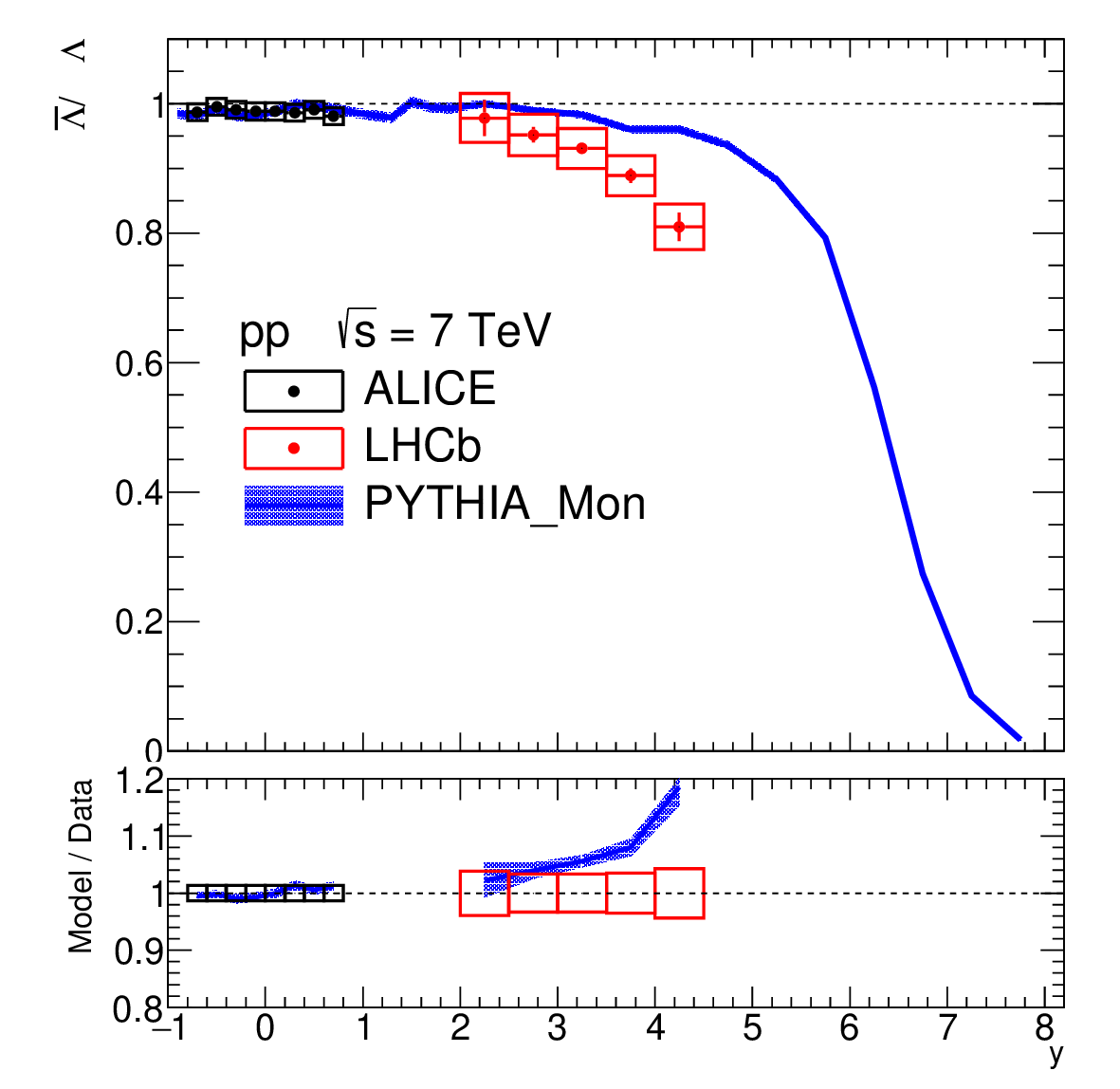}
    \label{fig:mon_comp}
  } 
  \hfill
  \caption{(a) Rapidity distribution of \lm and \antilm plotted using generated dataset with \namepythia 8.3 MC event generator
  in pp collisions at \seventev. The width of the blue patch of inset represents the statistical error of simulated data points. 
  \\ (b) $\bar{\Lambda}/ \Lambda$ ratio at \seventev as a function of rapidity. The black markers with black boxes 
  (systematic uncertainty) 
  represent ALICE collaboration's result ~\cite{alice2013mid} and the red markers with red boxes (systematic uncertainty) points show LHCb 
  collaboration's result\cite{aaij2011measurement}. The datapoints are compared with \pythiamode Mon dataset, the lower panel shows a comparison of $\bar{\Lambda}/ \Lambda$ ratio generated with \pythiamode Mon dataset to published data 
  of ALICE and LHCb collaborations.}
  \label{fig:mon_rapidity_comp}
\end{figure*}

The highest energy at which ALICE and LHCb collaborations' experimental results are available on  the
measurement of \antilm to \lm ratio is \tev $7$ TeV.
Hence for this investigation, events are generated using the \textsc{P}\textsc{ythia} 8.3 event generator at \tev $7$ TeV. Initially, $56\times 10^{6}$ pp events are generated with the
Monash (default) tune of \namepythia 8.3. Hereafter, for the sake of convenience, this dataset will be termed as `\pythiamode Mon'. Further, to stop the decay of the hadrons into their 
decay products, \textbf{HadronLevel:Decay} is turned off in the code. Using this dataset, the rapidity distributions of \lm and \antilm are plotted and shown in the upper panel 
of Fig.\ref{fig:mon_rapidity}. The ratio of \antilm to \lm as a function of rapidity, y ($ y = 0.5 \times ln [(E + p_{z})/(E - p_{z})]$) is shown in the lower panel of 
the same plot. 
From this plot, it can be readily seen that though the ratio is very close to unity at and around mid-rapidity, it certainly remains slightly less than 1.
Such an asymmetry in the production of baryons and their antibaryons can also be observed in 
relativistic heavy ion collisions due to the effect of nuclear stopping \cite{song2013baryon}. 
But, at LHC energies the proton-proton collisions are believed to be more or less transparent. Further, the asymmetry in the ratio increases with the increase of rapidity.
A comparison of this \pythiamode Mon prediction with experimental results of ALICE and LHCb is shown in Fig.\ref{fig:mon_comp}.
The upper panel of Fig.\ref{fig:mon_comp} shows rapidity dependent $\bar{\Lambda}/ \Lambda$ ratio at \seventev. The black markers with black boxes 
(systematic uncertainty) represent ALICE collaboration results ~\cite{alice2013mid} and the red markers with red boxes (systematic uncertainty) 
are the results of LHCb collaboration \cite{aaij2011measurement}. 

From 
Fig.\ref{fig:mon_comp}, it can be readily observed that while the \pythiamode Mon agreed well with the experimentally
measured \antilm to \lm ratio of ALICE collaboration at mid-rapidity, it fails to explain the observation from LHCb collaboration at large rapidity.
Here it is worth mentioning that  LHCb collaboration could not explain their experimental result with
\namepythia 6 LHCb MC, NOCR, and Perugia 0 tune generated MC data \cite{aaij2011measurement}.

\begin{figure*}
  \centering
  \subfigure[]{
    \includegraphics[width=0.45\linewidth]{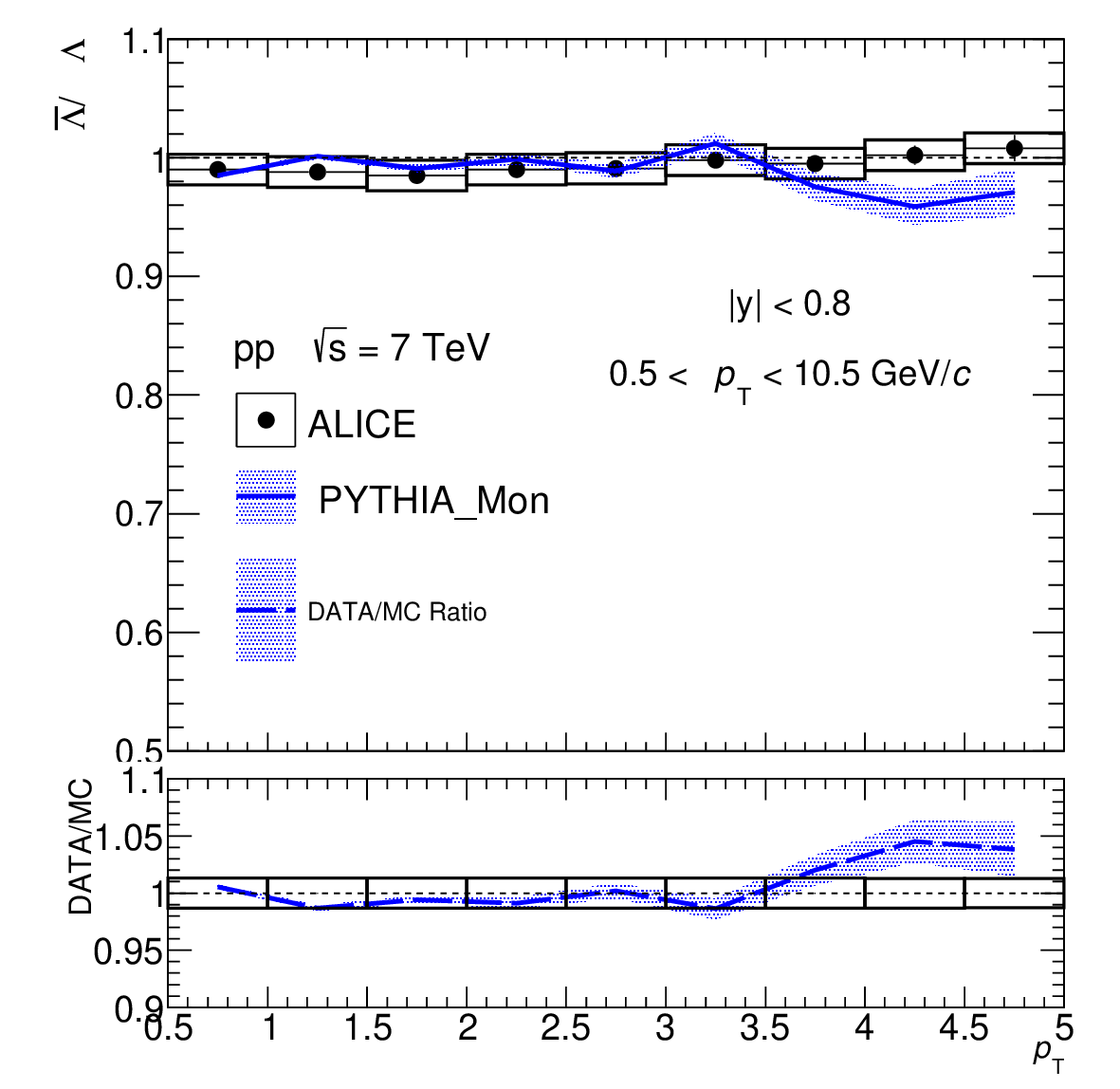}
    \label{fig:alice_ratio_pt_mon}
  }
  \subfigure[]{
    \includegraphics[width=0.45\linewidth]{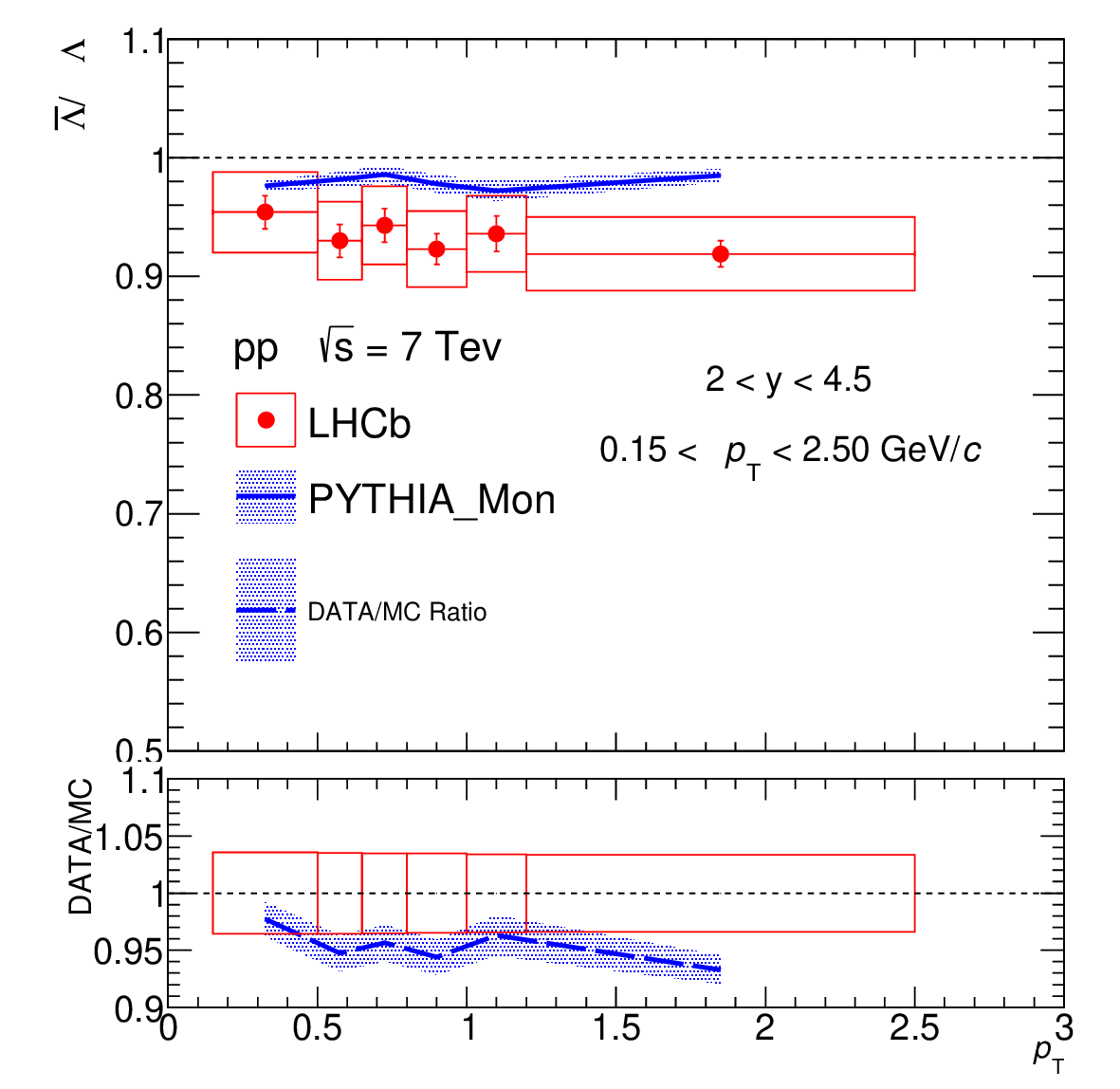}
    \label{fig:lhcb_ratio_pt_mon}
  } 
  \hfill
  \caption{$\bar{\Lambda}/ \Lambda$ ratio at \seventev as a function of \pt - (a) Within ALICE detector acceptance and (b) within LHCb detector acceptance. The black markers with black boxes 
    (systematic uncertainty) represent result of ALICE collaboration  ~\cite{alice2013mid} and the red markers with red boxes (systematic uncertainty) points show result of LHCb 
    collaboration \cite{aaij2011measurement}. The lower panel of both figures show, comparison of the datapoints from experiment 
    with simulated \pythiamode Mon dataset at the same energy.}
  \label{fig:pt_ratio_mon}
\end{figure*}

Fig. \ref{fig:alice_ratio_pt_mon} and \ref{fig:lhcb_ratio_pt_mon} show, $\bar{\Lambda}/ \Lambda$ ratio at \seventev as a function of transverse momentum, \pt
(\pt=$\sqrt{p_{x}^{2} + p_{y}^{2}}$). The black markers with black boxes (systematic uncertainty) represent ALICE collaboration results~\cite{alice2013mid} and the red markers points with 
red boxes (systematic uncertainty) show LHCb collaboration results\cite{aaij2011measurement}. From Fig. \ref{fig:alice_ratio_pt_mon}, it
can be observed that the ALICE experimental data points of the ratio of $\bar{\Lambda}/ \Lambda$ as a function of \pt has a tendency to exhibit values slightly less than 
unity particularly at low \pt region. From the lower panel of \ref{fig:alice_ratio_pt_mon}, 
it can be observed that the generated dataset \pythiamode Mon is more or less successful, well within the experimental uncertainty,  in describing the \pt dependent $\bar{\Lambda}/ \Lambda$ ratio of ALICE experimental result.
However, from Fig. \ref{fig:lhcb_ratio_pt_mon}, it can be observed that the same ratio measured by LHCb collaboration at a larger rapidity range ($2 < y < 4.5$), is much lower than unity. Again,
a comparison of the experimental ratio with \pythiamode Mon result, readily shows that 
\pythiamode Mon, here too, fails to predict the  $\bar{\Lambda}/ \Lambda$ ratio as measured by LHCb collaboration.
Here again, \namepythia 6 LHCb MC, NOCR, and Perugia 0 tunes were unable to explain LHCb collaboration's \pt dependent \antilm / \lm ratio for the full \pt range \cite{aaij2011measurement}. 
Such observations of $\antilm / \lm$ ratio different from unity even at mid rapidity indicates that pair production may not be the sole mechanism of 
\lm production in pp collision at this studied energy. It, therefore, becomes very much essential to look for a single model that includes better description of 
underlying pp collision dynamics  to explain the experimentally measured $\bar{\Lambda} / \Lambda$ ratio of ALICE and LHCb experiments (\& hence the enhanced \lm production).

\par
Therefore, in addition to dataset \pythiamode Mon, three more datasets are generated at the same energy by 
manipulating the parameters of \namepythia 8.3 Monash event generator. In one of these three sets, the 
effect of color reconnection is eliminated from the Monash tune via \textbf{ColourReconnection:reconnect=off}. This dataset is termed as `\pythiamode NOCR'. In another set, 
for adding the effect of string shoving to Monash tune, 
string shoving with \textbf{Ropewalk:doShoving = on} is turned on, and the flavour rope is turned off with \textbf{Ropewalk:doFlavour = off} \cite{bierlich2022comprehensive}.
This dataset is named as `\pythiamode SS'. At last, to study the effect 
of string junction, the formation of the junction is turned on via \textbf{BeamRemnants:beamJunction = on} which, otherwise, remains off in default version. 
This dataset is termed as `\pythiamode SJ'. Similar to \pythiamode Mon dataset, the decay of the hadrons into their decay products is turned off in all the above mentioned datasets. 
Event statics of various \namepythia 8.3 generated datasets are listed in Table \ref{tab:table}. 

\begin{table*}
  \caption{Events statistics of generated datasets using \namepythia 8.3 for pp collisions}
  \label{tab:table}      
  \begin{tabular}{ccccc}
   Center-of-mass & Physics modifications to& Dataset name& No. of events (in Millions)\\ 
   energy $\sqrt{s}$ (TeV) &  \namepythia event generator & &\\ \hline
   7& Used  default Monash tune & \pythiamode Mon &  56         \\
    & Included formation of string junction & \pythiamode SJ &  56         \\
    & Included string shoving mechanism& \pythiamode SS &  14        \\
    & Eliminated the effect of color reconnection & \pythiamode NOCR &  14         \\
   13.6&Included formation of string junction &\pythiamode SJ& 32\\
  \end{tabular}
\end{table*}

\par
\begin{figure}[!]
  \centering
  \includegraphics[width=0.9\linewidth]{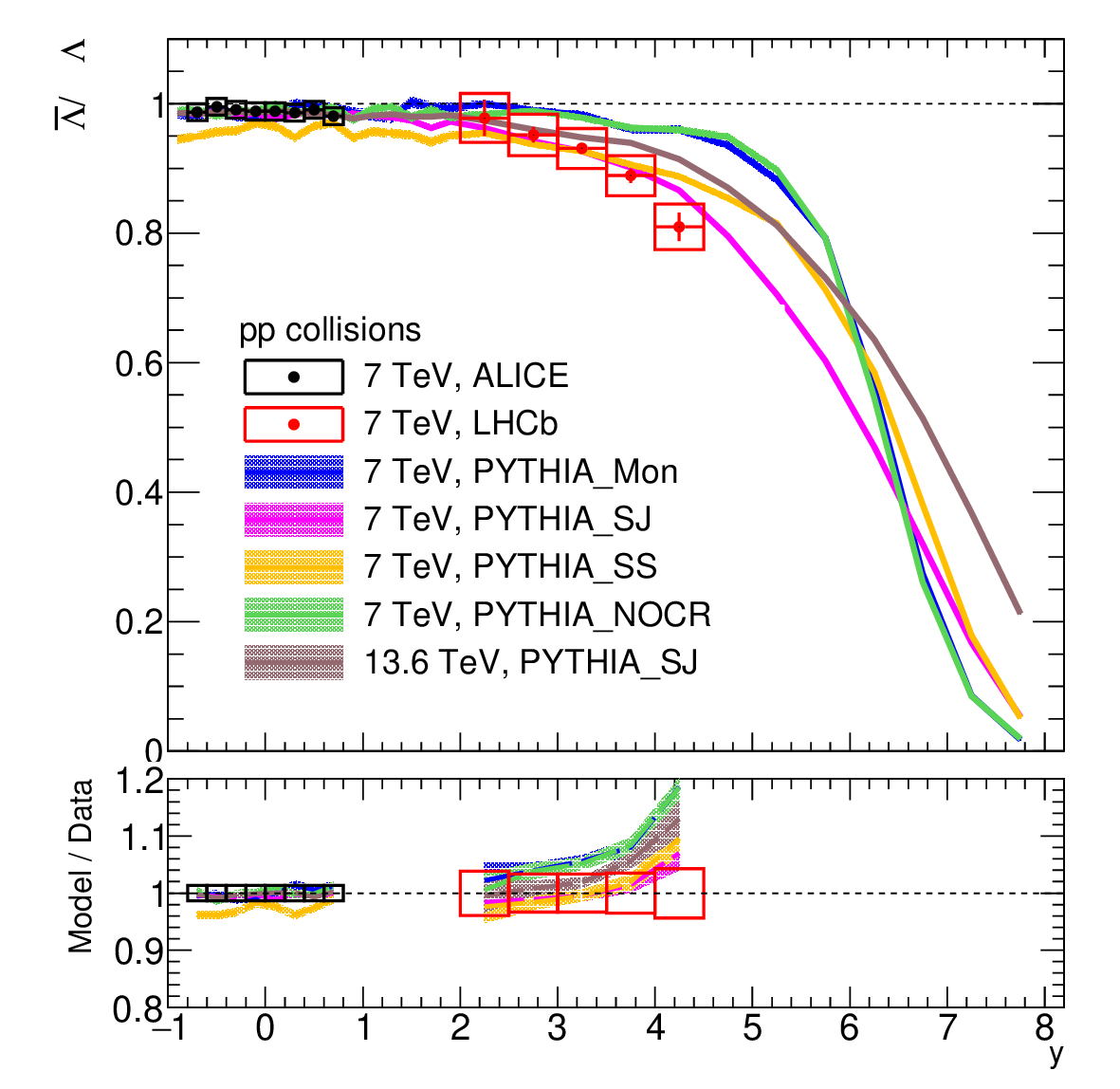}
  \caption{$\bar{\Lambda}/ \Lambda$ ratio at \seventev as a function of rapidity. The black markers with black boxes 
    (systematic uncertainty) 
    represent ALICE collaboration's result ~\cite{alice2013mid} and the red markers with red boxes (systematic uncertainty) points show LHCb 
    collaboration's result\cite{aaij2011measurement}. The datapoints are compared with various simulated datasets with 
    \textsc{P}\textsc{ythia} 8.3 event generator at \tev 7 and 13.6 TeV. The lower panel shows comparison of rapidity dependent $\bar{\Lambda}/ \Lambda$ ratio generated with various tune \namepythia 8.3 to published data 
    of ALICE and LHCb collaboration.}
  \label{fig:rapidity_data}
\end{figure}

With these MC datasets, the $\bar{\Lambda}/ \Lambda$ ratio as a function of rapidity is compared with experimental results of both ALICE and LHCb
for pp collisions at \tev $7$ TeV and is shown in Fig.\ref{fig:rapidity_data}. The lower panel of Fig.\ref{fig:rapidity_data}, shows the ratio
of $\bar{\Lambda}/ \Lambda$ using four generated datasets to both ALICE and LHCb experimental data. 
From Fig.\ref{fig:rapidity_data}, it can be readily seen that while the generated datasets of \pythiamode Mon, 
\pythiamode NOCR, \pythiamode SS fail to agree well with the experimental results of ALICE \& LHCb over the full rapidity space,
\pythiamode SJ, on the other hand, within the systematic uncertainty, agrees well with the experimental results of both the experiments over the full rapidity
space. Seeing the success of \pythiamode SJ in describing the rapidity dependent \antilm / \lm 
ratio over the ALICE and LHCb rapidity acceptance, we have generated a new set of data with 
\pythiamode SJ model for pp collision at 13.6 TeV (table \ref{tab:table}) to predict the the possible 
scenerio of baryon transportation in RUN3 pp data. It can be predicted that at this enhanced energy, baryon transportation will decrease, a result which support the experimental 
observation of ref. \cite{alice2013mid}.

\par

\begin{figure*}
  \centering
  \subfigure[]{
    \includegraphics[width=0.45\linewidth]{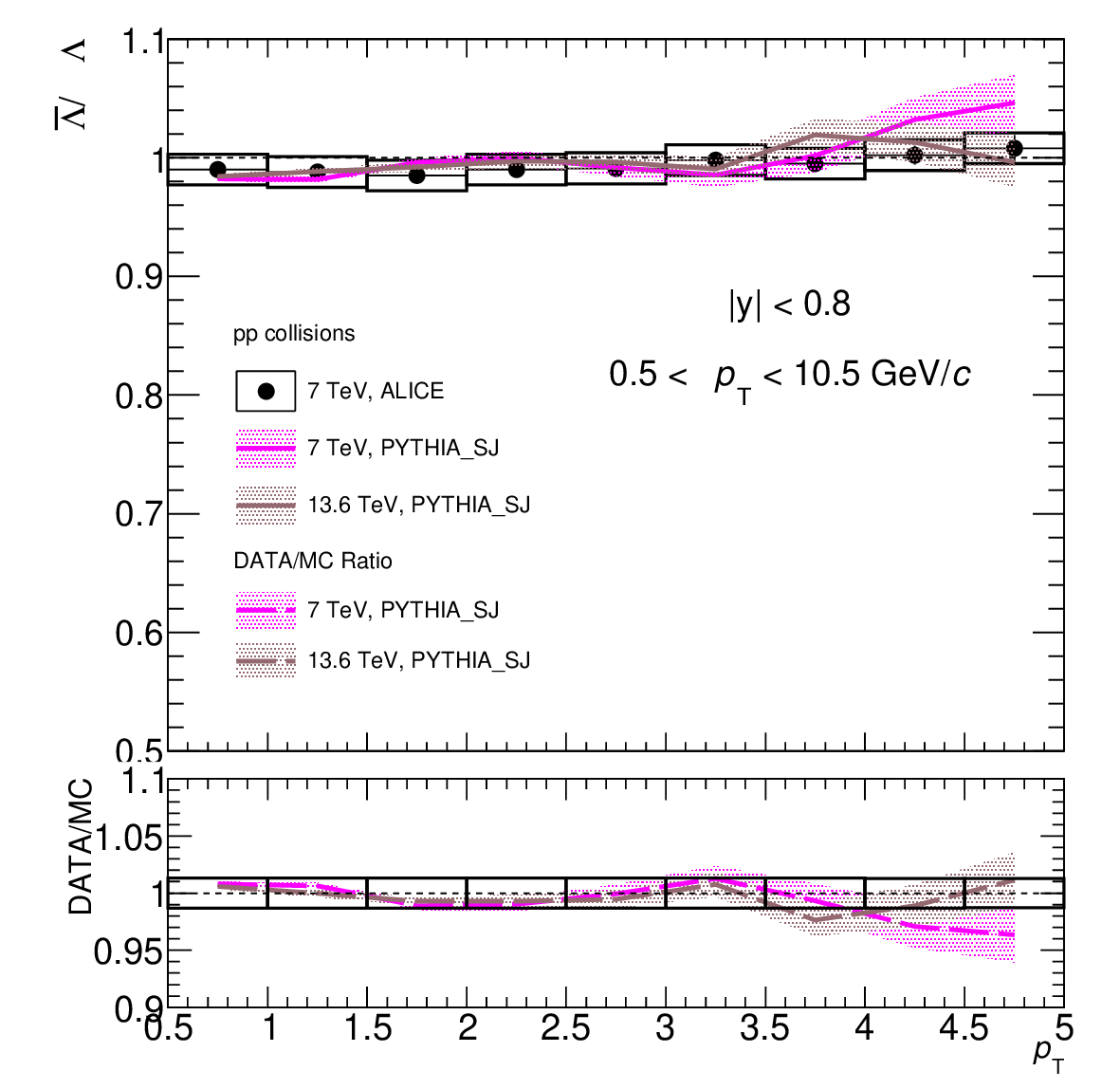}
    \label{fig:alice_ratio_pt}
  }
  \subfigure[]{
    \includegraphics[width=0.45\linewidth]{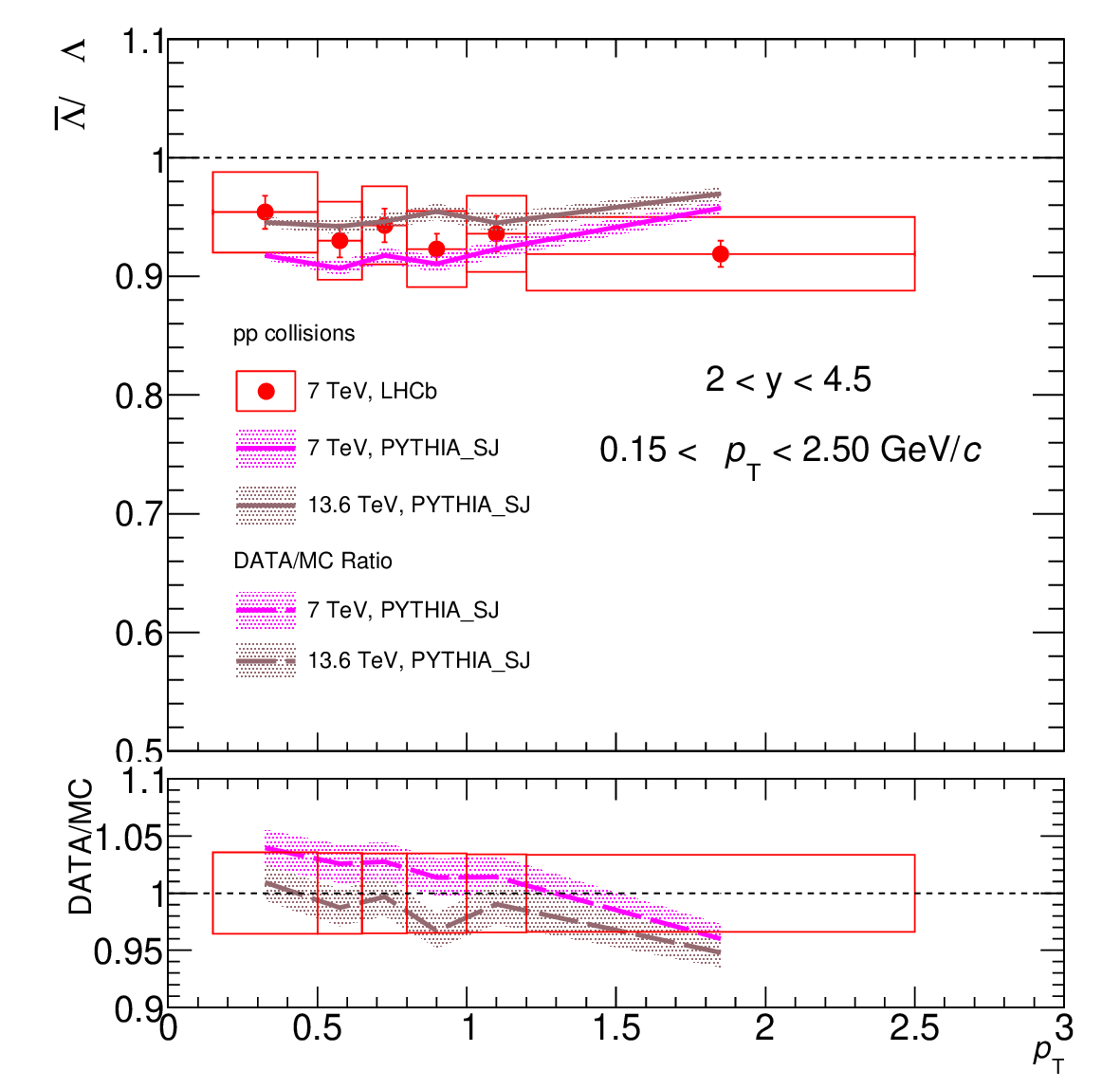}
    \label{fig:lhcb_ratio_pt}
  } 
  \hfill
  \caption{$\bar{\Lambda}/ \Lambda$ ratio at \seventev as a function of \pt - (a) Within ALICE detector acceptance and (b) within LHCb detector acceptance. The black markers with black boxes 
    (systematic uncertainty) 
    represent ALICE collaboration's result ~\cite{alice2013mid} and the red markers with red boxes (systematic uncertainty) points shows LHCb 
    collaboration's result\cite{aaij2011measurement}. The lower panel of both figures show, comparison of the datapoints from experiment 
    with simulated datasets using \pythiamode SJ at both \tev 7 and 13.6 TeV.}
  \label{fig:pt_ratio}
\end{figure*}

\begin{figure}[!]
  \includegraphics[width=0.9\linewidth]{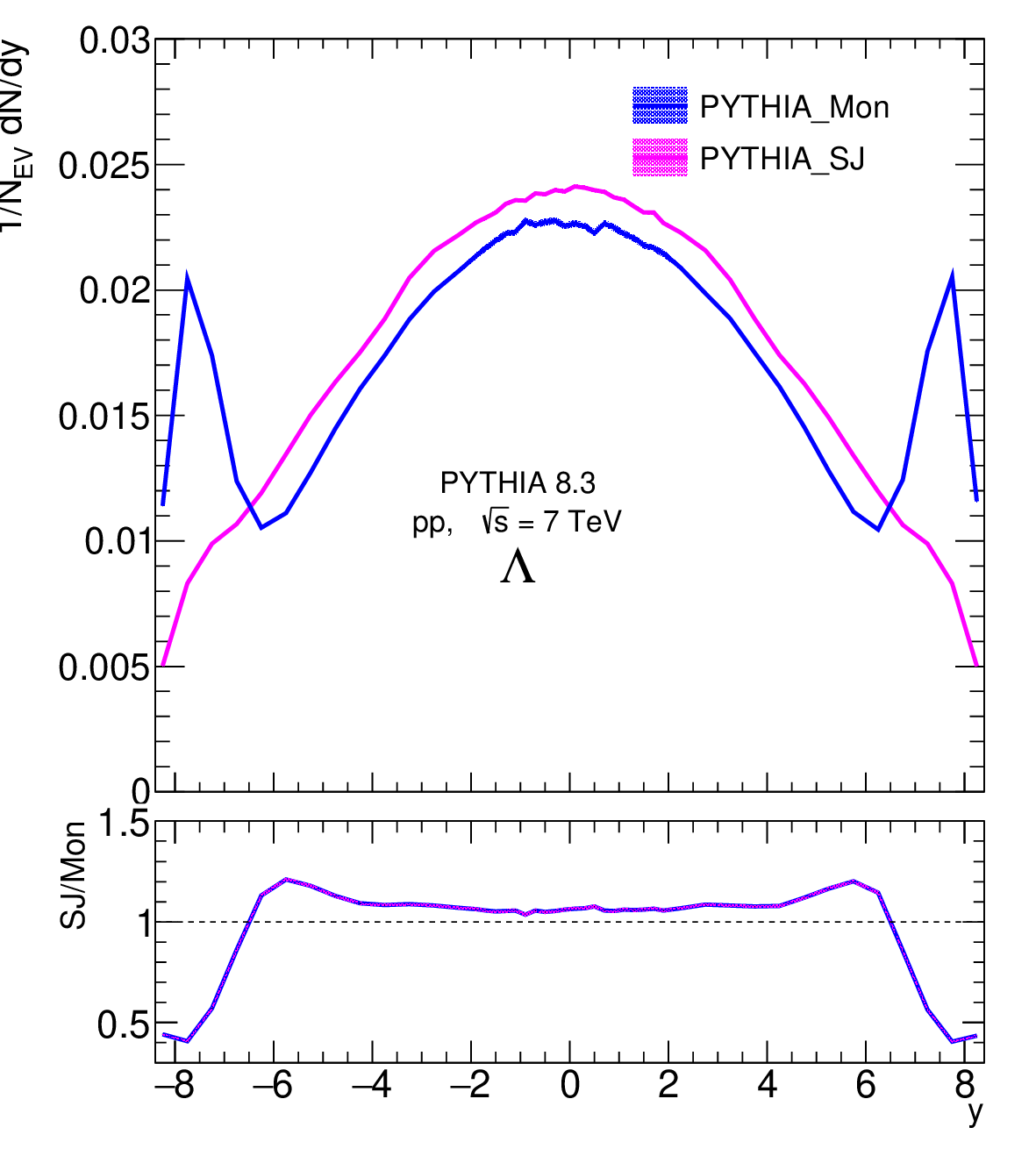}
  \caption{Rapidity distribution of \lm plotted using \pythiamode Mon (blue line) and \pythiamode SJ (pink line) simulated datasets
  in pp collisions at \seventev.}
  \label{fig:mon_vs_sj}
\end{figure}
Fig. \ref{fig:alice_ratio_pt} and \ref{fig:lhcb_ratio_pt} show the \pt spectra of \antilm/ \lm 
ratio drawn with ALICE and LHCb pp RUN2 data \cite{alice2013mid,aaij2011measurement} and our \pythiamode SJ generated data for the same system.
From Fig. \ref{fig:alice_ratio_pt} and \ref{fig:lhcb_ratio_pt}, it is readily evident that,
 within the experimental error,  the experimental data points of the spectra of both ALICE ($ -0.8 < y < +0.8$) and 
 LHCb ($2 < y < 4.5$) could well be reproduced with our generated \pythiamode SJ MC data set. As
 observed in Fig. \ref{fig:rapidity_data}, here too it can be expected that the asymmetry in the \pt spectrum 
 of \antilm / \lm ratio will decrease in pp RUN 3 data.

To have better understanding of the cause of asymmetry in \lm and \antilm production, with the \pythiamode SJ model generated data, the rapidity distribution of \lm is examined over the full rapidity range  
and compared with the result of \pythiamode Mon as shown in Fig. \ref{fig:mon_vs_sj}. The lower panel of the figure shows the ratio. \pythiamode Mon includes the quark-diquark fragmentation model and \pythiamode SJ incorporates 
junction topology in which, as a result, a string junction will be formed by breaking a diquark 
when more than two valence quarks are available in the beam remnant. 
As mentioned in ref \cite{kopeliovich1989novel}, 
the diquarks in the projectile fragmentation region are responsible for 
production of baryons in that region, which might have been the cause of 
the formation of two humps in the rapidity distribution of \lm in both 
fragmentation regions ($6 < \mody < 8$) as shown in \pythiamode Mon of 
Fig.\ref{fig:mon_vs_sj}. But \pythiamode Mon prediction does not match with the experimental results at LHCb acceptance. 
Considering diquark an indivisible
entity, it is not possible to move a baryon far from the beam fragmentation region to lower rapidity region \cite{kopeliovich1989novel}. So, Rossi and Veneziano introduced the 
concept of the color string junction topology (SJ)\cite{rossi1977possible, rossi1980theoretical}. 
With the introduction of SJ, the diquark will be destroyed, as a 
consequence, the baryon number can be transmitted across a large rapidity gap through SJ 
transmission from the fragmentation region. From the lower panel of Fig. \ref{fig:mon_vs_sj}, 
it is evident that with the inclusion of the string junction,  the transportation of baryon 
number from beam fragmentation towards LHCb rapidity acceptance region has been favoured 
and the two humps of the fragmentation  regions disappeared 
in \pythiamode SJ as the diquarks get destroyed.

\begin{figure}[!]
  \centering
  \includegraphics[width=0.9\linewidth]{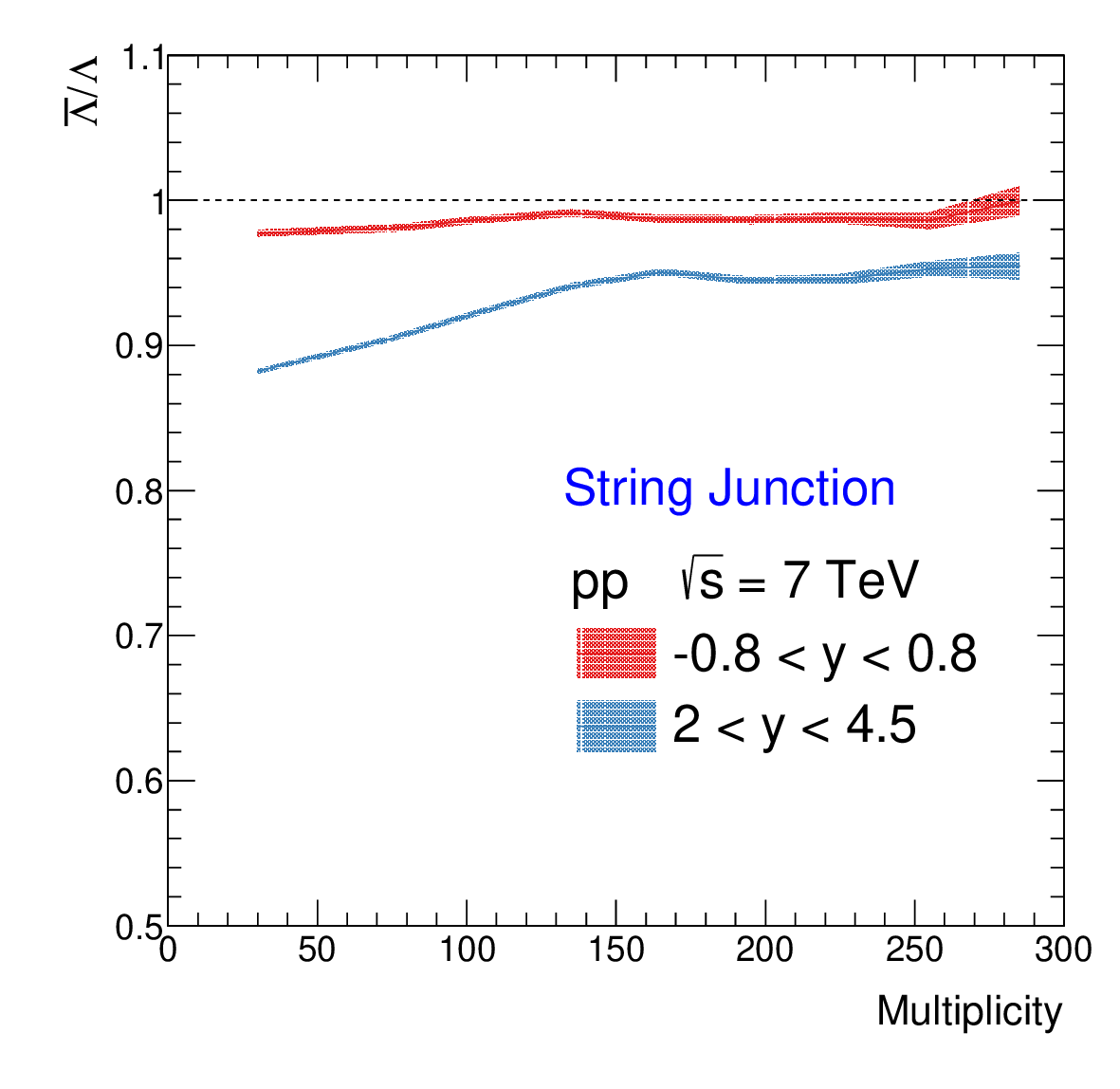}
  \caption{ $\bar{\Lambda}/ \Lambda$ ratio as a function of multiplicity in pp collision at \seventev with \pythiamode SJ generated data
   within ALICE (red) and LHCb (blue) rapidity acceptances.}
  \label{fig:multiplicity}
\end{figure}

As the high rapidity corresponds to beam fragmentation region, 
a study on multiplicity dependent $\bar{\Lambda}/ \Lambda$ ratio within both ALICE and LHCb detectors' acceptance is expected to shed more light on baryon number transportation.
Fig. \ref{fig:multiplicity} represents $\bar{\Lambda}/ \Lambda$ 
ratio as a function of multiplicity in pp collision at \seventev  with \pythiamode SJ data set within the acceptance of ALICE (red line) and LHCb (blue line)
detectors.
From this figure, it can be seen that while the ratio within the rapidity range \mody $< 0.8$ (ALICE acceptance) shows 
an increasing trend from low to high multiplicity, this increasing trend is much more faster in the LHCb acceptance $2 < y < 4.5$ (LHCb acceptance).
The observed increasing trend of  $\bar{\Lambda}/ \Lambda$ ratio with multiplicity, or otherwise, with centrality confirms that the baryon number 
indeed is transported from the beam fragmentation region. At the highest multiplicity class, the ratio approaches unity at both ALICE and LHCb acceptances.

\begin{figure*}
  \centering
  \subfigure[]{
    \includegraphics[width=0.45\linewidth]{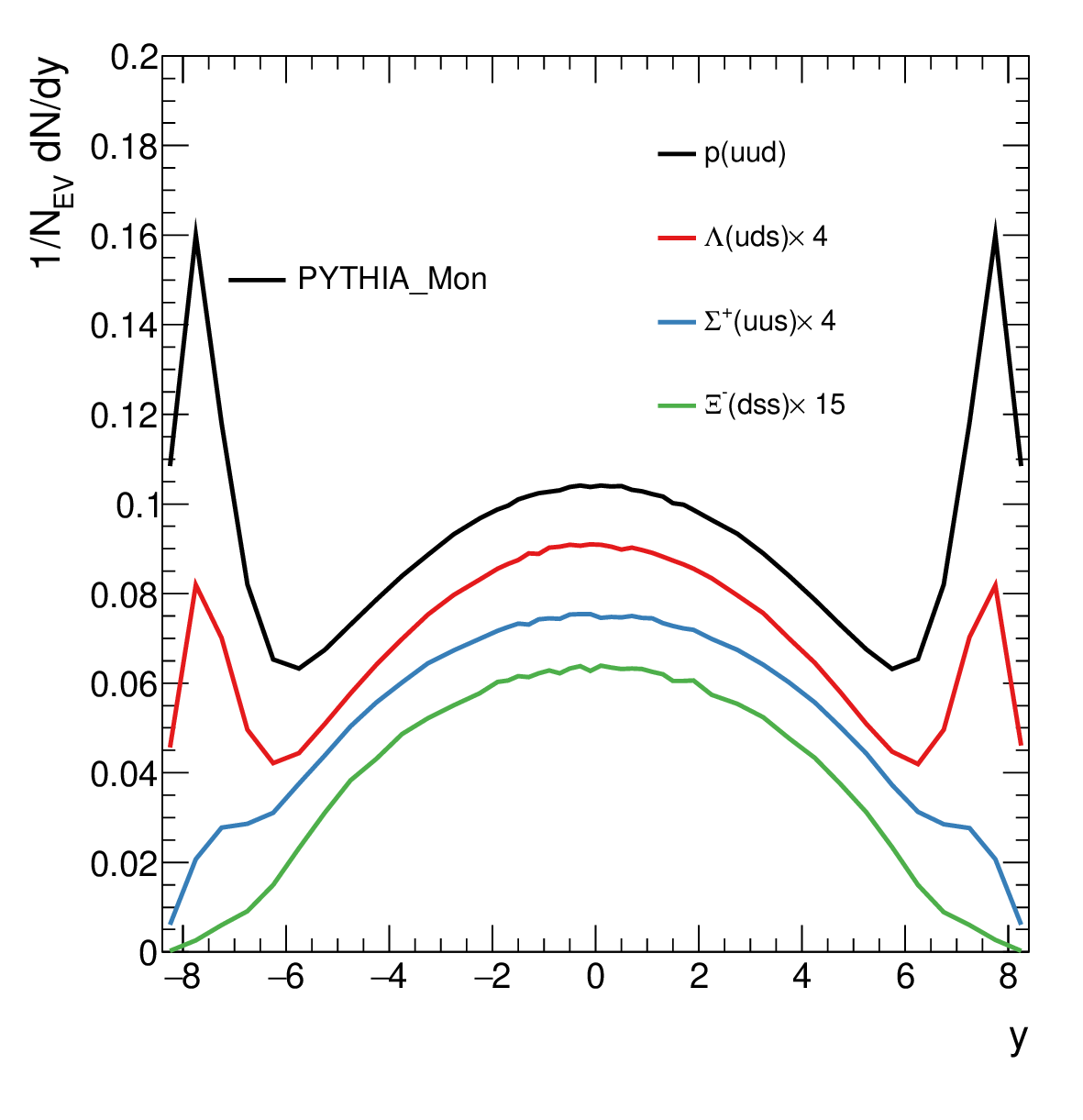}
    \label{fig:rapidity_bar_mon}
  }
  \subfigure[]{
    \includegraphics[width=0.45\linewidth]{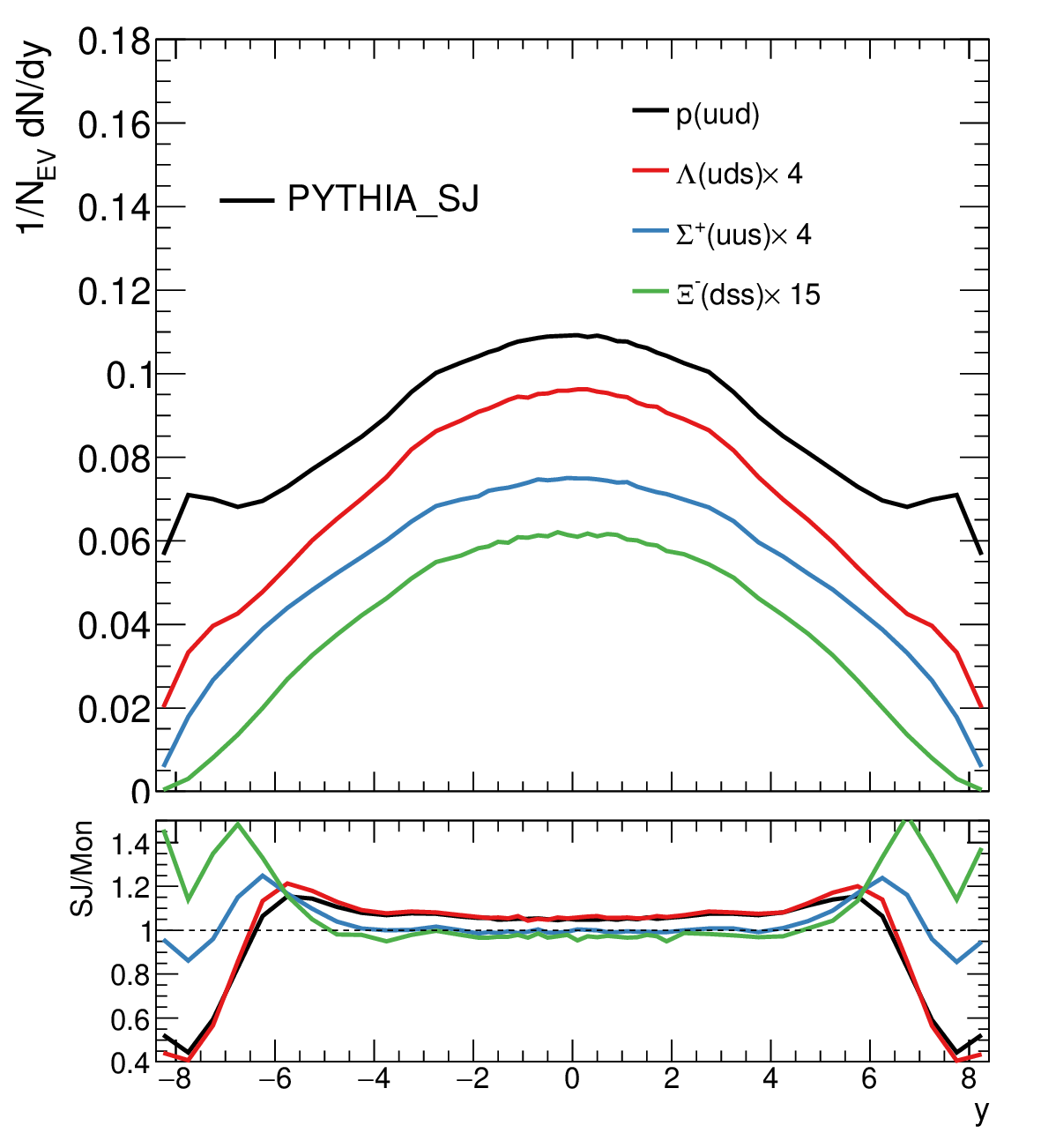}
    \label{fig:rapidity_bar_sj}
  } 
  \hfill
  \caption{Rapidity distribution of proton and hyperons plotted with - (a) \pythiamode Mon (b) \pythiamode SJ in pp collisions at \seventev}
  \label{fig: hyperons}
\end{figure*}
\begin{figure*}[!]
  \centering
  \includegraphics[width=0.9\linewidth]{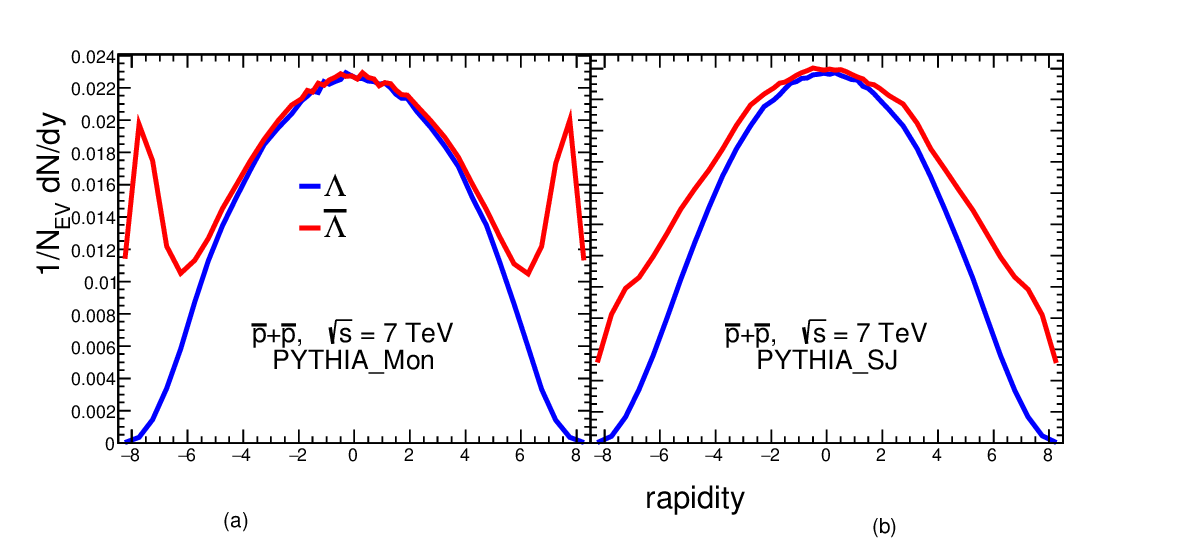}
  \caption{Rapidity distribution \lm and \antilm  in $\bar{p} + \bar{p}$ collision at \seventev with (a) \pythiamode Mon, (b) \pythiamode SJ generated data}
  \label{fig:apap}
\end{figure*}  
Fig.\ref{fig: hyperons}(a) and (b) show the rapidity distribution of protons and hyperons that 
include at least one leading (valance) quark (u/d) with \pythiamode Mon and \namepythia SJ respectively. It can be readily seen that the yield of baryons in 
the beam fragmentation region decreases with the decrease in the number of valance (u/d) quark content of 
a baryon for both Monash and SJ-generated data. Moreover, those baryons that are produced more in number 
in the beam fragmentation region (or otherwise the baryons that contain more u/d quarks) are transported 
more to the central rapidity region making SJ/Monash ratio greater than unity as shown in the lower panel of
Fig.\ref{fig: hyperons}(b). Such an observation clearly indicates that the baryon number transportation indeed 
took place from the beam fragmentation region to the mid-rapidity region with the fragmentation of 
diquarks and the formation of string junctions. Thus, these plots supplement our earlier claim of 
easy transportation (throwing) of baryon number from beam to mid-rapidity region with the formation of 
string junction.

For further insight into the fact that baryon number really transport from the fragmentation region towards 
mid rapidity region, we have generated a new set of data with both Monash and SJ PYTHIA models at \seventev for $\bar{p} + \bar{p}$ collision 
and plotted again the rapidity distribution of \lm and \antilm. It is expected that for $\bar{p} + \bar{p}$ collision, 
the beam fragmentation region should be populated with anti-string junction rather than string-junction. 
In such a situation, the production of anti-baryon (anti-lambda) should be more in the extreme rapidity 
region than the baryon (\lm) and the transportation of anti-baryon number, rather than baryon number, should take place. 
From the Fig.\ref{fig:apap}, we can see that the generated data corroborate our expectations and claims of transportation of (anti-) baryon number from
the fragmentation region towards the central rapidity region.

\section{Summary}

In this work, the \namepythia 8.3 event generator with various tunes (default, NOCR, SS \& SJ) is used to study the \(\bar{\Lambda}/\Lambda\) 
ratio as a function of rapidity (\(y\)), transverse momentum (\(p_T\)), and multiplicity and compared with the experimentally measured \(\bar{\Lambda}/\Lambda\) ratio of the ALICE and LHCb collaborations. 
While the \pythiamode SJ data set shows quite a good agreement with the experimental results of 
both ALICE and LHCb, the other data set generated with default, NOCR, and SS tunes couldn't.
Additionally, the experimentally measured \(\bar{\Lambda}/\Lambda\) ratio as a function of \(p_T\) also shows good 
agreement with the \pythiamode SJ dataset  over the full rapidity of ALICE and LHCb acceptance. The \pythiamode SJ includes 
the formation of string junctions among the partons from beam remnant. These string junctions that carry the 
baryon numbers are believed to be transported from beam rapidity to mid-rapidity region. Such transportation of baryon number is supposed to be cause of observation of $\antilm / \lm $ less than unity in 
both ALICE and LHCb experiment at \tev 7 TeV for pp system. A multiplicity dependent study of $\antilm / \lm $ ratio vindicated such transportation of baryon number
through string junction. Baryon number transportation is expected to decrease in LHC RUN3 pp data. For $\bar{p} + \bar{p}$ collisions generated data at \seventev, where there is more anti-baryon in the fragmentation region, the anti-baryon number is found to be transported towards the central rapidity region making \antilm / \lm greater than unity, a scenario just opposite to p+p collisions.

\section{ACKNOWLEDGMENTS}
The authors thankfully extend their gratitude to the \namepythia8.3
 team for developing the codes and generously 
making them freely available to the public domain. Additionally,
the authors gratefully acknowledge the financial support provided by the Department of Science and Technology 
(DST), Government of India, under project No. $SR/MF/PS-02/2021-GU(E-37122)$ to carry out this work and 
providing JRF fellowship to B. Barman.

\end{document}